\documentclass[pra,twocolumn,superscriptaddress,showpacs]{revtex4}
\usepackage[english]{babel}

\usepackage{graphicx}
\newcommand{\lf}{\left}
\newcommand{\rg}{\right}
\newcommand{\be}{\begin{equation}}
\newcommand{\ee}{\end{equation}}
\newcommand{\bea}{\begin{eqnarray}}
\newcommand{\eea}{\end{eqnarray}}
\newcommand{\ba}{\begin{array}}
\newcommand{\ea}{\end{array}}
\newcommand{\bd}{\begin{displaymath}}
\newcommand{\ed}{\end{displaymath}}

\newcommand{\g}{\gamma}
\newcommand{\G}{\Gamma}
\newcommand{\D}{\Delta}
\renewcommand{\d}{\delta}

\newcommand{\s}{\sigma}
\newcommand{\vf}{\varphi}
\newcommand{\ve}{\varepsilon}

\renewcommand{\o}{\omega}

\renewcommand{\k}{\kappa}
\newcommand{\nn}{\nonumber}
\newcommand{\ra}{\rangle}
\newcommand{\la}{\langle}
\newcommand{\hf}{\frac{1}{2}}
\begin{document}

\title{Phase diffusion and locking in single-qubit lasers}
\author{Stephan Andr\'e}
\affiliation {Institut f\"ur Theoretische Festk\"orperphysik,
       Universit\"at Karlsruhe, 76128 Karlsruhe, Germany}
\author{Valentina Brosco}
\affiliation {Institut f\"ur Theoretische Festk\"orperphysik,
       Universit\"at Karlsruhe, 76128 Karlsruhe, Germany}
\affiliation{Dipartimento di Fisica, Universit\`a ``La Sapienza'', 
P.le A. Moro 2, 00185 Roma, Italy}       
\author{Alexander Shnirman}
\affiliation{Institut f\"{u}r Theorie der Kondensierten Materie,
Universit\"{a}t Karlsruhe, 76128 Karlsruhe, Germany}
\affiliation {DFG Center for Functional Nanostructures (CFN),
       Universit\"at Karlsruhe, 76128 Karlsruhe, Germany}
\author{Gerd Sch\"on}
\affiliation {Institut f\"ur Theoretische Festk\"orperphysik,
       Universit\"at Karlsruhe, 76128 Karlsruhe, Germany}
\affiliation {DFG Center for Functional Nanostructures (CFN),
       Universit\"at Karlsruhe, 76128 Karlsruhe, Germany}

\begin{abstract}
Motivated by recent experiments, which demonstrated lasing and 
cooling of the electromagnetic field in an electrical resonator 
coupled to a superconducting qubit, we study the phase coherence and 
diffusion of the system in the lasing state. We also  discuss  phase 
locking and synchronization induced by an additional {\sl ac} 
driving of the resonator.   We extend earlier work to account for 
the strong qubit-resonator coupling and to include the effects of 
low-frequency qubit's noise. We show that the strong coupling may 
lead to  a double peak structure of the spectrum, while the shape 
and width are determined to the low-frequency noise. 
\end{abstract}

\pacs{}

\maketitle

In a number of recent experiments (here we only cite few examples) 
superconducting qubits 
coupled on chip to electrical or mechanical resonators displayed 
quantum electrodynamic effects and opened the field of ``circuit 
QED" 
\cite{wallraff04,ilichev03,grajcar,johansson,naik,schuster,astafiev,chiorescu,sillanpaa}. 
By creating a population inversion between two charge 
states in a driven superconducting single-electron transistor (SSET) 
Astafiev {\sl et al}.~\cite{astafiev} demonstrated lasing behaviour of a microstripline resonator coupled to the qubit. 
Grajcar {\sl et al}.~\cite{grajcar}  coupled a 
driven flux qubit to a low-frequency  $LC$ resonator and observed both 
cooling and a trend towards lasing of the resonator field.  
 In contrast to usual lasers, where many atoms are weakly 
coupled to the electromagnetic field, in single-qubit lasers one 
 artificial atom is coupled strongly to the resonator. 
In addition,  solid state qubits are subject to decoherence effects.
Some of the consequences and novel behavior
 had been analyzed in Refs. 
\cite{blais04,rodrigues07-1,you,hauss08,zhirov,marthaler08}. 

Even in the lasing state, the coherence of the electromagnetic field 
is lost due to 
phase diffusion after a characteristic time $\tau_d$ \cite{haken},  
an effect which is observable, e.g.,  in the laser spectrum. 
Phase diffusion can be suppressed by \emph{injection 
locking}, that is by driving the resonator with an additional 
coherent signal. This fixes the phase difference between the laser 
and  driving field to a value which depends on the intensity and 
 detuning of the latter. Both injection locking and phase 
diffusion were studied experimentally for a single-qubit laser in 
Ref.~\cite{astafiev}.  As compared to the spectrum observed in 
standard (many-atom) lasers, the single-qubit laser spectrum  is 
broader and the peak is substantially shifted with respect to the 
natural resonator frequency. The maximum photon number in the 
resonator is rather low, raising questions about the coherent nature 
of the amplification in these systems. 

In this work we study the spectral properties of single-qubit lasers 
and explain qualitatively several of the experimental observations. 
In Section \ref{model} we introduce the model and describe our 
approach. Results for static properties of single qubit lasers are 
discussed in Section \ref{static}. Here we focus on the average 
photon number in the vicinity of the lasing transition, which 
illustrates the differences between  single- and multi-atom lasers. 
We show explicitly that single qubit lasers are characterized by  a 
smooth transition to the lasing regime and by the absence of a sharp 
lasing threshold. Next, in Section \ref{diff}, we analyze the phase 
diffusion of the resonator field focusing on the following issues: 
(i) we discuss the effects of correlations between qubit and 
resonator on the diffusion process; (ii) we show how the interplay 
between strong coupling and spontaneous emission  may lead to a 
double peak structure in the spectrum; and (iii) we demonstrate how 
low-frequency noise leads to inhomogeneous broadening of the lasing 
peak. Finally, in Section \ref{injection}, we study injection 
locking induced by an external coherent driving, we discuss the main 
features of the spectrum, and we provide an estimate for the locking 
threshold.  

\section{Model} \label{model} We consider a single-mode quantum resonator  coupled to 
$N_a$ qubits  (labelled by $\mu$) and we account for  both  
resonator and qubit dissipation.  In the rotating wave approximation 
the system is described  by the  Hamiltonian
\bea\label{H} H & =& \! \hbar \o_0 a^{\dagger} a +\frac{1}{2}\hbar 
\o_p\sum_{\mu}\s^{\mu}_z 
+ \hbar g\sum_{\mu} \lf( \s^{\mu}_+ a + \s^{\mu}_- a^{\dagger}\rg) \\
\nonumber
& +&(a + a^{\dagger}) X_a +\sum_{\mu}\lf(X^{\mu}_z \s^{\mu}_z + X^{\mu}_{+} 
\s^{\mu}_+ + X^{\mu}_{-} \s^{\mu}_- \rg) +H_{\rm N}. \eea
Apart from the photon operators, $a$ and $a^{\dag}$, we introduced  
the Pauli matrices acting on the single-qubit eigenstates 
$\s^{\mu}_z=\lf|1_{\mu}\rg>\lf<1_{\mu}\rg|-\lf|0_{\mu}\rg>\lf<0_{\mu}\rg|$, 
$\s^{\mu}_{+}=\lf|1_{\mu}\rg>\lf<0_{\mu}\rg|$, 
$\s^{\mu}_{-}=\lf|0_{\mu}\rg>\lf<1_{\mu}\rg|$. Dissipation is 
modelled by assuming that the oscillator and  the qubits interact 
with noise operators, $X_a$ and $X^{\mu}_z$, $X^{\mu}_+$, 
$X^{\mu}_{-}$, belonging to independent baths with Hamiltonian 
$H_{\rm N}$ in thermal equilibrium. The noise coupling 
longitudinally to the qubits,  $X^{\mu}_z \s^{\mu}_z$, is 
responsible for the qubits' pure dephasing. To describe lasing, we 
assume that a population inversion has been created in the qubits, 
which we describe by assuming that the effective temperature of the 
qubit baths is negative.  In this way (for $N_a=1$) we model the 
essential properties of the 
SSET laser used by Astafiev {\sl et 
al}.~\cite{astafiev,marthaler08}. Possible deviations from the 
standard Jaynes-Cummings  oscillator-qubit coupling used in Eq. 
(\ref{H}) are discussed in Appendix A. From the Hamiltonian 
(\ref{H}), following the route described, e.g., in 
Ref.~\cite{haken}, we derive a set of quantum Langevin equations of 
motion,
\bea
& & \frac{d}{dt} \sigma^{\mu}_z = - 2 i g \lf( \s^{\mu}_{+} a - \s^{\mu}_{-} a^{\dagger} \rg)  - \G_1 (\sigma^{\mu}_z - D_0) + F^{\mu}_z(t) \nn\ ,\\
& & \frac{d}{dt}\sigma^{\mu}_+ = - \lf(\G_{\vf}-i\o_p\rg) 
\sigma^{\mu}_+ - i g
\s^{\mu}_z a^{\dagger} + F^{\mu}_+(t)\nn\ ,\\
& & \frac{d}{dt} a = -\lf(\frac{\k}{2}+i \o_0\rg) a - i g \sum_{\mu}
\s^{\mu}_{-} +
F_a(t).\label{heis3} \eea
Here the rate $\G_1=\Gamma_{\downarrow}+\Gamma_{\uparrow}$ is the 
sum of excitation and relaxation rates, while 
$\G_\vf=\Gamma_1/2+\G_{\vf}^*$ is  the total dephasing rate, which 
also accounts for pure dephasing due to the longitudinal noise 
described by $\G_{\vf}^*$. In contrast to relaxation and excitation 
processes, $\G_{\vf}^*$ accounts for processes with no energy 
exchange between qubit and environment, which thus do not affect the 
populations of the two qubit states.  Furthermore, $\k$ is the bare 
resonator damping.  The parameter $D_0 = 
(\Gamma_{\uparrow}-\Gamma_{\downarrow})/\Gamma_1$ denotes the 
stationary qubit magnetization in the absence of the resonator. In 
the present case, since we assume a negative temperature of the 
qubits baths and a population inversion we have $D_0>0$. The 
Langevin operators  $F^{\mu}_i(t)$ with $i=+,-,z$ have vanishing 
averages and are characterized by their correlation functions, $ \la 
F^{\mu}_{i}(t)F^{\nu}_{j}(t')\ra = 
\d_{\mu\nu}D^{\mu}_{ij}g_q(t-t')$. The function $g_q(t-t')$ is 
assumed to decay on a time scale much shorter than the relaxation 
and decoherence times of the qubits and the oscillator. The 
diffusion coefficients $D^\mu_{ij}$ are related to the rates 
introduced above, 
$D^{\mu}_{+-}=\Gamma_{\uparrow}+\G_{\vf}^*(1+\la\s^{\mu}_z\ra)$,  
$D^{\mu}_{-+}=\Gamma_{\downarrow}+\G_{\vf}^*(1-\la\s^{\mu}_z\ra)$,  
$D^{\mu}_{zz}= 2 \Gamma_1 - 2 
(\Gamma_{\uparrow}-\Gamma_{\downarrow})\la\s^{\mu}_z\ra $, 
$D^{\mu}_{z+}=2 \Gamma_{\downarrow}\la\s^{\mu}_{+}\ra$, 
$D^{\mu}_{z-}=- 2 \Gamma_{\uparrow}\la\s^{\mu}_{-}\ra$.  Similarly, 
the Langevin force $F_a(t)$ acting on the resonator with $\la 
F^{\dag}_a(t)F_a(t')\ra = \k N_{\rm th} g_a(t-t')$ is characterized 
by the rate $\k$ and thermal photon number $N_{\rm th}$. The qubit 
and oscillator noises are assumed to be independent. For a further 
discussion of the different rates and diffusion coefficients we 
refer to standard quantum optics textbooks, e.g. Ref. \cite{cohen}.
 
\section{Static properties} \label{static}
\begin{figure}
\includegraphics[width=8cm]
{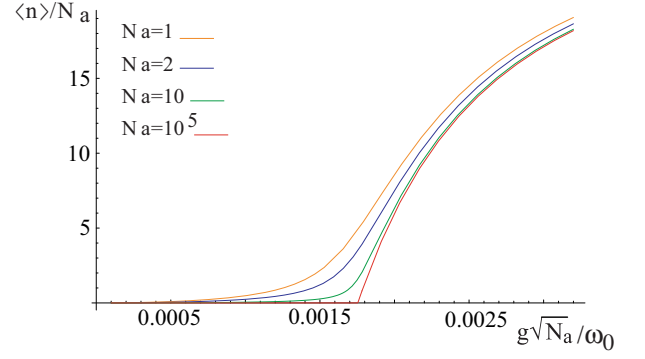} \vskip 5mm\caption{Color online - Scaled photon 
number, $\la n\ra/N_a$, in the threshold region versus scaled 
coupling, $g\sqrt{N_a}$ for different values of $N_a$. The other 
parameters are: $\o_p = \o_0$, $\G_1/\o_0 = 0.016$, $\G_\vf^*/\o_0 = 
0.004$, $D_0 = 0.975$, $\k/\o_0 = 3 \cdot 10^{-4}$ and $N_{th} = 0$. 
} \label{transition}
\end{figure}

By taking appropriate products of Eqs. (\ref{heis3}) and performing 
averaging, we arrive at the following equations for the  average 
photon number, $\la n\ra$, the qubit polarization $\la\s^{\mu}_z\ra$ 
and the product $\la\sigma^{\mu}_+ a\ra$:
\begin{eqnarray}
   & & \frac{d}{dt} \langle \s_z^\mu\ra=- 2 i g \lf( \la\s^{\mu}_{+} a\ra - \la\s^{\mu}_{-} a^{\dagger}\ra \rg)  - \G_1 (\la\sigma^{\mu}_z\ra - 
   D_0)\\
   & & \frac{d}{dt}\langle  n \rangle = i g \sum_{\mu} \left( \langle \sigma^{\mu}_+ a \rangle - 
   \langle \sigma^{\mu}_- a^{\dagger} \rangle \right) - \kappa \left( \langle n \rangle - N_{th} \right), \label{eqn}\\
   & & \frac{d}{dt} \langle \sigma^{\mu}_+ a \rangle = \left(  i \Delta-\gamma  \right) 
   \langle \sigma^{\mu}_+ a \rangle - i g \langle \sigma^{\mu}_z n \rangle - i \sum_{\nu} \langle \sigma^{\mu}_+ \sigma^{\nu}_- \rangle, 
   \nn\\\label{eqs+a}
\end{eqnarray}
where we introduced the detuning $\D=\o_p-\o_0$ and the total 
dephasing rate $\g=\Gamma_\vf+\frac{\k}{2}$.

 In the stationary limit, neglecting the correlations between 
different qubits, i.e. assuming $\langle \sigma^{\mu}_+ 
\sigma^{\nu}_- \rangle \simeq \delta_{\mu\nu}(1+\langle 
\sigma_z^{\mu} \rangle)$, the previous equations yield the following 
two exact relations   between three quantities,  the average 
polarization  $\la S_z(t)\ra$ with $S_z \equiv 
\frac{1}{N_a}\sum_{\mu}\s_z^{\mu}$, the photon  number  $\la n(t) 
\ra$,  and  the correlation function $\la n S_z \ra$, 
 \bea
     \la n \ra & =& N_{th} + \frac{2 g^2 N_a}{\k} \frac{\g}{\g^2 + \D^2}
\lf[ \la S_z n \ra + \frac{\la S_z \ra + 1}{2}\rg],  \, \nn\\
     \la S_z \ra & = & D_0 - \frac{4 g^2}{\G_1} \frac{\g}{\g^2 + \D^2}
 \lf[ \la S_z n \ra + \frac{\la S_z \ra + 1}{2}\rg]\,.\label{stat}
\eea 
If one of them is known, e.g., from a numerical solution of the 
Master equation, the other two can readily be determined. 

 Factorizing the 
correlator, $\la S_z n \ra \approx \la S_z \ra\la  n \ra$, on the 
right-hand side of Eqs.~(\ref{stat}) gives results known in quantum 
optics as ``semi-quantum model" \cite{mandel}. It  includes 
spontaneous emission processes, described  by the term proportional 
to $(\la S_z\ra+1)$. \\
Spontaneous emission has a twofold importance for the issues 
described in the present work. First, at low temperatures it is 
responsible for the line-width of the lasers. Second, as noticed in 
Ref. \cite{mu}, due to the low photon number, spontaneous emission 
is especially relevant in the description of the dynamics of single 
atom lasers.  To illustrate this fact, we plot  in 
Fig.~\ref{transition} the scaled photon number $\la n\ra/N_a$  as a 
function of the scaled coupling $g\sqrt{N_a}$ for different values 
of $N_a$. The product $g\sqrt{N_a}$ is kept constant to have a 
universal asymptotic behaviour.  In the limit of large $N_a$ we 
observe a sharp lasing transition occurring at the threshold 
coupling $g_{\rm thr}\sqrt{ N_a}= \sqrt{\kappa\gamma / (2 D_0)}$, as 
predicted by the semiclassical theory. On the other hand, for low 
values of $N_a$, and in particular for $N_a=1$, we find  a smooth 
crossover  between the normal and the lasing regimes, which is due 
to spontaneous emission. In this case, although we cannot define a 
sharp threshold condition, we can still identify a transition region  
centered at the semiclassical threshold coupling.\\The results 
presented in Fig. 1 were obtained analytically using  the 
semi-quantum approximation. In the case $N_a=1$ we compared such 
results with the numerical solution of the Master Equation and we 
obtained an agreement better than $10^{-3}$ . 

\section{Phase diffusion} \label{diff}For typical circuit QED parameters, i.e., 
for strong coupling $g$, the semi-quantum approximation, in spite of 
giving, as explained above, a good estimate of the stationary photon 
number,  cannot be used to study spectral functions. For the 
analysis of phase diffusion we 
 thus proceed with a \emph{hybrid} approach: starting from 
Heisenberg equations of motion we derive analytical expressions for 
the phase correlation time and  frequency shift of the lasing 
peak, both expressed as functions of the single-time averages, i.e., the photon number and qubit 
inversion in the stationary state. We then use the Master equation 
for the reduced qubit-resonator density matrix to calculate 
the single-time averages. 

\begin{figure}
\includegraphics[width=7cm]
{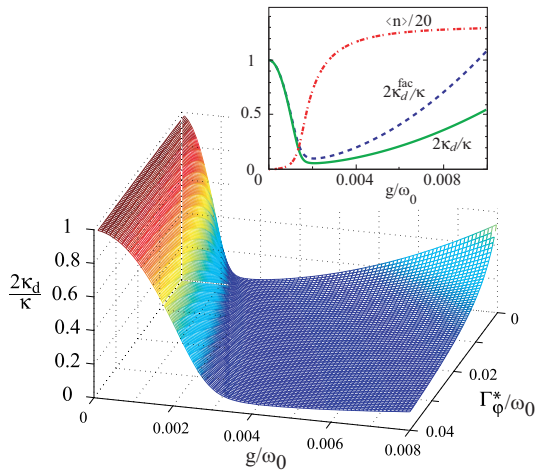} \vskip 5mm
\caption{Color online - Diffusion constant versus qubit-oscillator 
coupling and pure dephasing; parameters as in Fig.\ref{transition}. 
Inset: phase diffusion constants, $\k_d$  and $\k_d^{\rm fac}$, 
calculated with  and without taking into account correlation and the 
average photon number obtained from the Master equation for  
$\G_{\vf}^*/\o_0 = 0.004$. } \label{comp}
\end{figure}

From now on we consider a single-qubit laser, $N_a=1$,
and analyze  the laser and  cross  correlation functions
\be \ba{rcl} O(\tau)&=&\lim_{t\to \infty}\la 
a^\dag(t+\tau)a(t)\ra,\\[0.2 cm] G(\tau)&=&\lim_{t\to \infty}\la 
\s_{+}(t+\tau)a(t)\ra.\ea\ee
Starting from the quantum Langevin equations (\ref{heis3}) we derive 
a hierarchy of equations involving $O(\tau)$ and $G(\tau)$. To 
truncate the hierarchy we split $a(t)$ into an  amplitude and phase, $a(t) = \sqrt{n(t)+1}\,e^{-i\varphi(t)}$, and assume that 
the correlation time of  phase fluctuations, $1/\k_d$, is much 
longer than that of  amplitude fluctuations, $\sim 1/\k$ 
\cite{haken}. This allows us to approximate for sufficiently long 
times, $\tau > 1/\k$, 
\be\la\s_z(t+\tau)a^\dag(t+\tau)a(t)\ra\simeq\frac{\la\s_z 
\sqrt{n}\ra}{\la \sqrt{n}\ra}\la 
a^\dag(t+\tau)a(t)\ra,\label{approx}\ee  while the correlator 
$\la\s_z \sqrt{n}\ra$  can be estimated as 
\be \frac{\la\s_z \sqrt{n}\ra}{\la \sqrt{n}\ra} \simeq
\frac{1}{2}\left(\la\s_z\ra  + \frac{\la\s_z n\ra}{\la n\ra}\right)
. \ee 
Above threshold this approximation, which neglects terms of 
order $\la\d n^2\ra/\la n\ra^2$, is justified when 
the fluctuations of the photon number are much smaller 
than the average.   
Starting from Eqs.~(\ref{heis3}) and 
using  the factorization (\ref{approx}) we  
obtain a coupled set of equations,
\be \ba{rcl} \frac{d}{d\tau}O(\tau)&=& \lf(i \o_0- 
\frac{\k}{2}\rg)O(\tau)+ i g G(\tau)\ ,\\[0.2cm] 
\frac{d}{d\tau}G(\tau)&=& \lf(i\o_p-\G_{\vf}\rg)G(\tau)- i g 
\frac{\la\s_z \sqrt{n}\ra}{\la \sqrt{n}\ra} O(\tau)\ . 
\ea\label{c1}\ee
These equations depend implicitly on all the parameters specified 
above.   We focus on the case where the oscillator damping is much 
weaker than qubit's  dephasing, $\k/2 \ll \G_{\vf}$, which is 
usually satisfied in single-qubit lasing experiments. In this  case 
we obtain from Eqs.~(\ref{c1}) for the oscillator's  spectral 
function, i.e., the real part of the Fourier transform of the 
correlator $O(\tau)$, a Lorentzian,  
%
\be   \hat O(\o)= \frac{2 \k_d \la n \ra}{(\o-\o_0-\d\o_0)^2+\k_d^2} 
\, . \label{spec} \ee 
It depends on the phase diffusion rate 
\bea\k_d&=&\frac{\k}{2}\frac{N_{th}}{\la 
n\ra}+\frac{g^2\G_{\vf}}{2\la n\ra}\frac{\lf( \la \s_z \ra + 1 
\rg)}{\G_{\vf}^2+\D^2}\nn\\ & & 
+\frac{g^2\G_{\vf}}{\G_{\vf}^2+\D^2}\frac{\la\s_z n\ra-\la\s_z\ra\la 
n\ra}{2\la n\ra} \label{kd} \, ,\eea
and the frequency shift \bea \d\o_0 &=&  \frac{\D}{2 \la n 
\ra}\lf[\frac{\k\lf(\la 
n\ra-N_{th}\rg)}{\G_{\vf}}-g^2\frac{\la\s_z\ra 
+1}{\G_{\vf}^2+\D^2}\rg]\nn\\ & 
&-\frac{g^2\D}{\G_{\vf}^2+\D^2}\frac{\la\s_z n\ra-\la\s_z\ra\la 
n\ra}{2\la n\ra} \, . \label{fs}\eea
The spectral function $\hat O(\o)$ is proportional to the spectrum 
measured in Ref. \cite{astafiev}.
 Equations (\ref{kd}) and (\ref{fs}) are the main 
results of the present work. Upon factorization of the correlator 
$\la\s_z n\ra$ far above and below the lasing transition they reduce  
to results known from quantum optics \cite{haken}.  The phase 
diffusion rate (\ref{kd}) is the sum of three terms. The first 
represents a thermal  contribution to the linewidth and is 
negligible in the experimental regime explored in Ref. 
\cite{astafiev}. The second  describes  the effect of 
 relaxation processes of the qubit, which, due to the 
strong coupling, strongly increases the  linewidth. The third term, 
due to quantum correlations, is essentially a measure of the 
coherent coupling between the qubit and the oscillator and  leads  
to a {\sl reduction} of the linewidth. Its effect is illustrated in 
the inset of  Fig.~\ref{comp}, where  we plot the diffusion constant 
$\k_d$ given by (\ref{kd}) covering the whole range from below to 
above threshold and compare it to the diffusion rate, $\k_d^{\rm 
fac}$, obtained by factorizing the correlator $\la\s_z n\ra$. As one 
can see correlations give significant quantitative corrections.  
Furthermore, we note that there is a wide range of parameters in 
which our approximations remain consistent; in the case of strong 
coupling $g$  this corresponds indeed to have  $\k_d<\k$ (see 
discussion around Eq.~(\ref{approx})).

 Fig.~\ref{comp} also displays the dependence 
of the phase diffusion rate on the pure Markovian dephasing rate 
$\Gamma_{\vf}^*$, showing a reduction of the linewidth with 
increasing $\G_{\vf}^*$ above threshold. This surprising feature is 
a consequence of the fact that pure dephasing processes are not 
associated with emission of incoherent photons in the resonator, and 
 their main effect is simply a  decrease of the 
effective qubit-resonator coupling. As one notes from the inset 
of Fig.~\ref{comp}, in single-qubit 
lasers far above threshold a reduction of the coupling has little 
effect on the saturated photon number but leads to a decrease of  
the incoherent photon emission rate, thus diminishing the 
linewidth. 

Another interesting manifestation of this effect is shown in the 
left panel of Fig. \ref{spectrum}. Here we plot the height of the 
spectral line $\hat O(\o_0+\d\o_0)=\la n\ra/\k_d$ as a function of 
the detuning, $\D$. One can see that the optimal lasing conditions 
are realized  somewhat out of resonance, where the effective 
coupling is weaker. We thus observe two peaks in the spectrum 
symmetrically shifted with respect to $\D=0$. Due to the strong 
coupling the photon number is roughly constant in the region between 
the two peaks. A similar structure in the  output  spectrum  of 
single atom laser was also found in a numerical study by Ginzel 
\emph{et al.} \cite{briegel}. One might conjecture that this effect 
is the origin of the two spots observed in the experiment. It would  
explain why the peaks  do not occur at resonance, but we have not 
succeeded to fit the experimental data in a satisfactory way
\footnote{In Ref. \cite{astafiev}, Astafiev \emph{et al.} study the 
resonator spectrum as a function of the charging energy of the SSET 
and observe two bright spots, blue and red detuned with respect to 
the $\D=0$ condition. We remark, that since in the experiment the 
Josephson energy is kept constant the detuning depends monotonically 
on the charging energy. Moreover a variation of the charging energy 
not only changes the detuning but also the coupling and the 
inversion. This does not change substantially the structure of the 
spectrum, but it leads to asymmetries between two peaks.}. In Ref. 
\cite{astafiev} it was proposed that the second peak is related to 
two-photon processes. Indeed deviations from the 
 model used in Eq. (\ref{H}), lead to an 
effective two photon-coupling between the qubit and the resonator. 
However, as described in the appendix, the two-photon coupling 
constant seems to be too small to produce any ``two-photon lasing''.
This results is also confirmed by the numerical solution of the 
Master Equation.  We also note that in the experiment the second 
peak appears substantially shifted from the two-photon resonance 
condition, using the data of Ref. \cite{astafiev}, at the second 
peak  we have   
$2\o_0-\o_p\simeq 0.4\o_0 $.\\

 The linewidth of order of 0.3MHz observed in  Ref.~\cite{astafiev} is 
about one order of magnitude  larger  than what follows from  
Eq.~(\ref{kd}) (of the order of the Schawlow-Townes linewidth). 
Moreover in the experiment the laser line shows a Gaussian rather 
than a Lorentzian shape. Both  discrepancies may be explained if we 
note that the qubit's dephasing is mostly due to low-frequency 
charge 
noise, which 
cannot be treated within the Markov approximation used in the 
derivation of Eqs.(\ref{spec})-(\ref{fs}). However, low-frequency 
(quasi-static) noise can be taken into account by averaging the 
Lorentzian line in Eq. (\ref{spec}) over different detunings 
\cite{falci}. Assuming that the detuning fluctuations are  Gaussian 
distributed  with mean $\bar \D$  and width $\sigma$,   such that 
$\Gamma_1>\sigma\gg \k_d$, we can neglect in the saturated limit the 
fluctuations of $\k_d$ and $\la n \ra$   and assume that the 
frequency shift $\d\o_0$ depends linearly on the detuning $\Delta$. 
From Eq. (\ref{fs}) we then have $\d\o_0\simeq 
\D\k/(2\Gamma_{\vf})$, and we  obtain a Gaussian line with width 
$\tilde\s \simeq \s\k/(2\Gamma_{\vf})$, where we remark that 
$\Gamma_\vf$ is the total markovian dephasing rate.   The linewidth 
observed in the experiment is then reproduced by a reasonable value 
of $\s$ of order of 300 MHz. In the case in which $\sigma$ is larger 
than $\Gamma_1$,  the previous formula overestimate the linewidth 
since it doesn't take into account the decay of $\la n\ra$ out of 
threshold. In this case one can perform the averaging numerically. 
Anyway, in the presence of low-frequency noise, the linewidth is 
governed not by $\k_d$ as one may have expected, but by $\d\o_0$.

\section{Injection locking} \label{injection} We next investigate 
the behavior of the single-qubit laser when the oscillator is driven 
by an external laser field or seed light. To describe  monochromatic 
driving with frequency $\o_{\rm dr}$ and amplitude $E_0$, we add a 
term $E_0 a e^{i \o_{\rm dr} t} + {\rm H.c.}$ to the Hamiltonian 
(\ref{H}). It leaves  the equations for the qubit operators 
unchanged, but modifies the quantum Langevin equation for the 
resonator,  
\be \frac{d}{dt} a = -\lf(i\o_0 +\frac{\k}{2} \rg) a - i E_0^* e^{-i
\o_{\rm dr} t} - i g \s_{-} + F_a(t) \, .\label{heis4} \ee
The average $\la a\ra$ now acquires a non-vanishing value and 
oscillates with the driving frequency $\o_{\rm dr}$. In the resonant 
case $\o_{\rm dr} = \o_0 = \o_p$, we estimate $\la 
\tilde{a}^{\dagger} \ra \equiv \la ae^{+ i \o_{\rm dr} t} \ra = -i 
E_0/\bar\k_d$, where $\bar 
\k_d=\hf\lf(\k_d+\sqrt{(\k_d^2+2|E_0|^2/\la n\ra}\rg)$ and  $\k_d$ 
has the same functional dependence on $\la n\ra$ and $\la\s_z\ra$ as 
in the  undriven resonant case,  
$\k_d=\frac{\k}{4}\lf(1+\frac{N_{th}}{\la n \ra}\rg)+\frac{g^2}{2 
\Gamma_\vf}\lf(\frac{\la\s_z\ra+1}{2\la n\ra}-\la \s_z\ra\rg)$.

%
%
\begin{figure}[t!]
\begin{minipage}[b]{4cm}
\centering
\includegraphics[width=3.8cm]{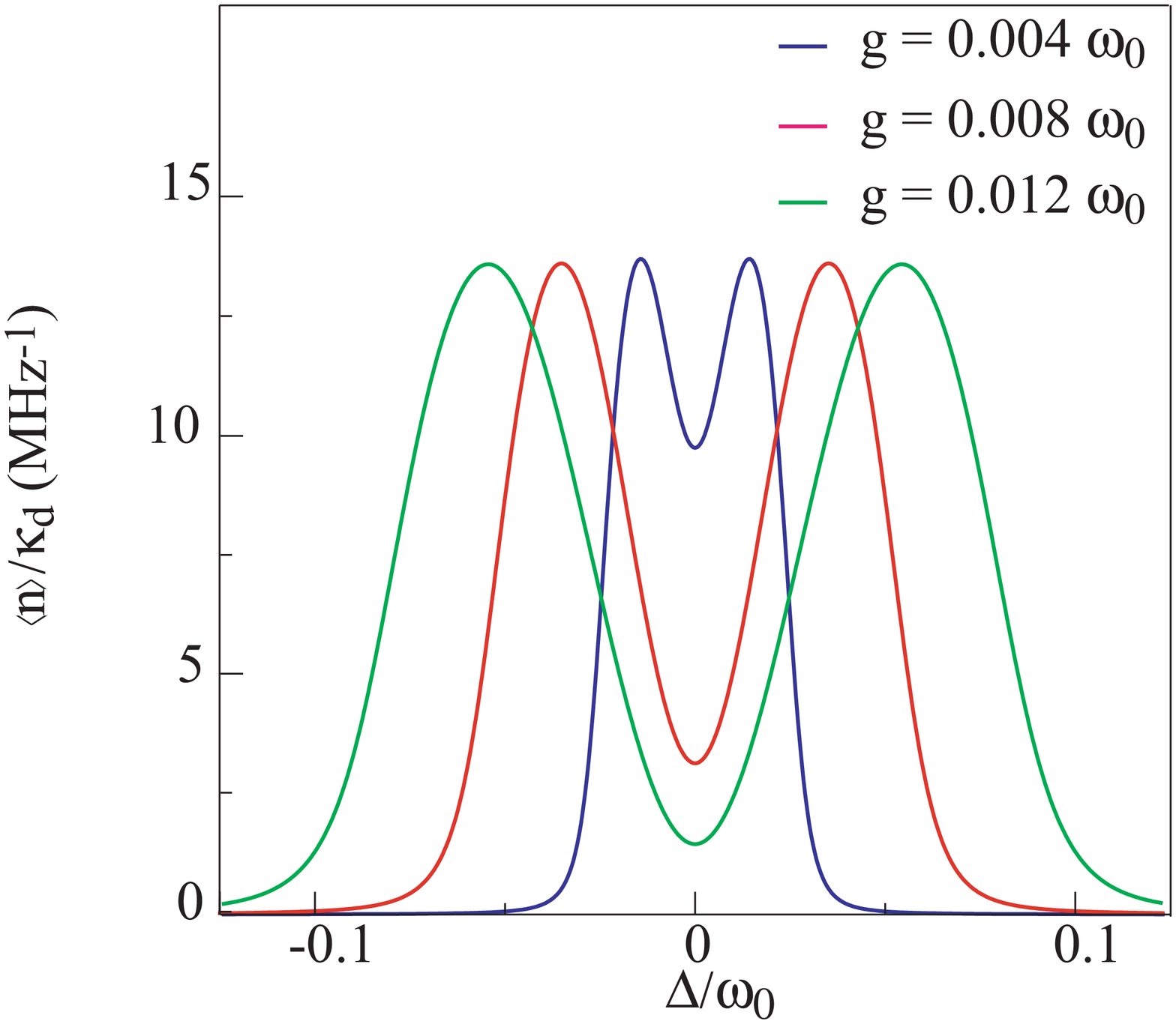}
\end{minipage}
 \hspace{3mm} 
\begin{minipage}[b]{4cm}
\centering
\includegraphics[width=5.5cm]{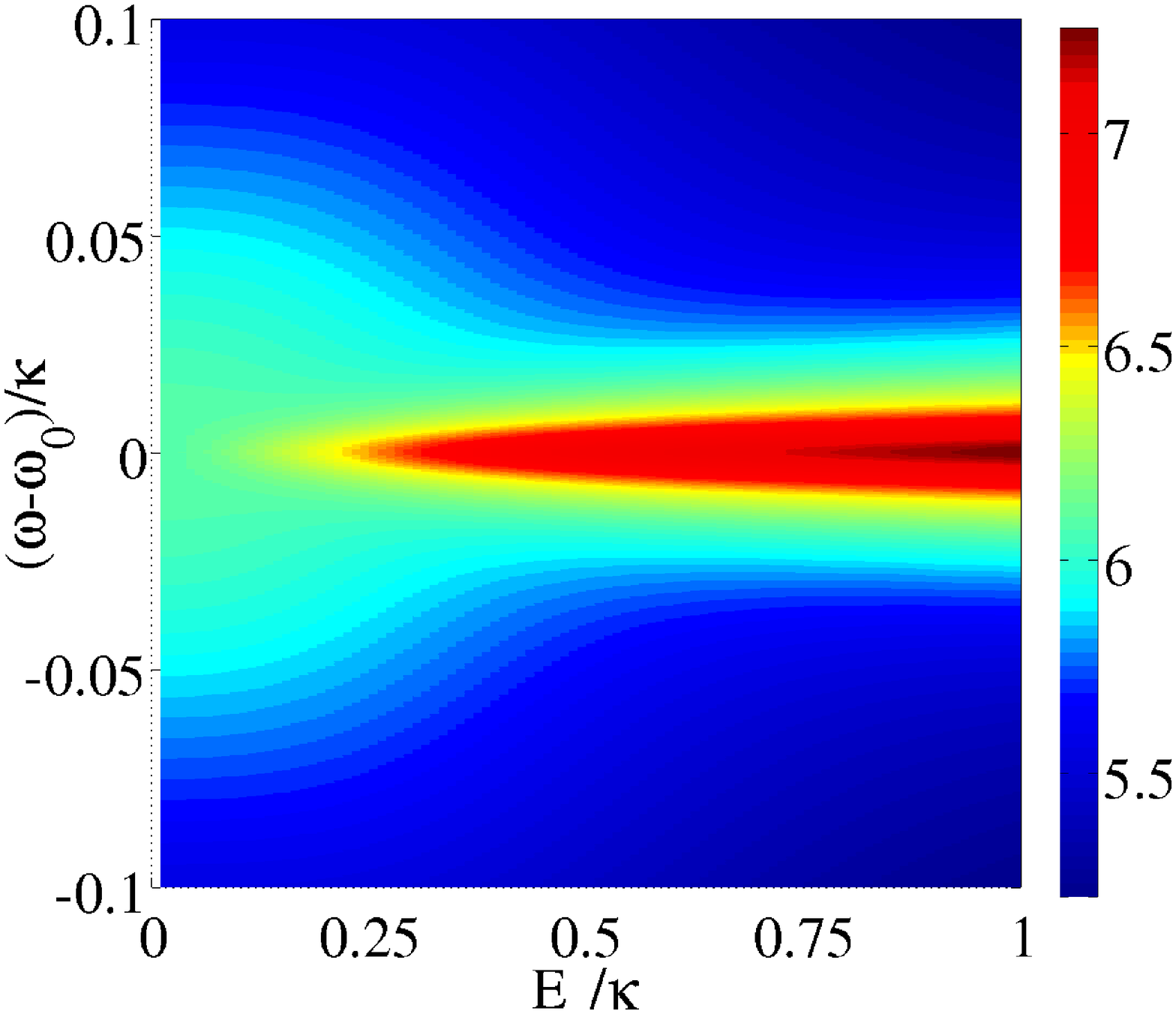}
\end{minipage}
\caption{Color online - Left panel: Maximum spectrum's amplitude  
$\hat O(\o_0+\d\o_0)$ for different values of $g$.  Right panel: 
Logarithm of the normalized resonator spectrum, $\log[\o_0\hat 
O(\o)]$ in  the presence of an external driving at resonance as a 
function of the driving power. $\o_0=10 \mathrm{GHz}$ other 
parameters as in Fig.~\ref{transition}.}\label{spectrum}
\end{figure}

Finally, we consider the emission spectrum of the single qubit laser 
in the injection locking regime. For simplicity we neglect the 
low-frequency noise. In the double resonance regime, that is for 
$\o_{\rm dr} = \o_0 = \o_p$, we get a simple  analytical expressions 
for $\hat O(\o)$,
\be\label{inlock} \hat O(\o) = \frac{2\bar \k_d \lf( \la n \ra - \la 
\tilde{a}^{\dagger} \ra \la \tilde{a} \ra \rg)}{(\o-\o_0)^2+\bar 
\k_d^2} + 2\pi \d(\o-\o_0) \la \tilde{a}^{\dagger} \ra \la \tilde{a} 
\ra \, . \ee
For low driving amplitude $E_0$, the resonator output is thus the 
superposition of two signals: the Lorentzian lasing peak and a 
coherent peak due to the driving proportional to $\lf|\la \tilde 
a\ra\rg|^2$.  As one  can see by combining the previous equation  
with the expression of $\la \tilde a^{\dag}\ra$, with increasing  
$E_0$, the height of the Lorentzian decreases and approaches zero, 
while the coherent peak grows. Eventually for large values of $E_0$, 
only the latter, which is amplified due to the coupling to the 
qubit, is visible in the spectrum. The driving amplitude, $\bar 
E_0$, at which the Lorentzian peak disappears can be evaluated, 
using Eq.(\ref{inlock}),  as $\bar E_0^2\simeq\la n\ra \bar \k^2_d$. 
The comparison to the experimental results of Ref. \cite{astafiev} 
can be done estimating the mean power absorbed by the system, $P_d$,  
as follows: $P_d\simeq E_0^2\o_0/\bar\k_d$ \cite{cohen}. 
 To illustrate 
the locking  transition we plot in Fig. \ref{spectrum} (right panel) 
the spectrum (\ref{inlock}). The output spectrum is centered at 
$\o=0$ since $\d\o_0=0$. In the numerics we assumed that the 
injected signal has a Lorentzian shape with width $w_{\rm 
dr}=\k/200$. In the presence of a finite qubit-oscillator detuning 
$\D = \o_p - \o_0$, the position of the lasing peak would be 
shifted. \section{ Conclusion}  We analyzed in detail the spectral 
properties of single-qubit lasers. Our main conclusions are:\\
 - Due to the strong coupling nontrivial structures appear 
in the spectrum, which are not visible in the average photon number. 
As shown in Fig. \ref{spectrum} the optimal lasing 
conditions are  realized for two values of the detuning, which are symmetrically shifted from $\D=0$. At 
these two hot-spots the output spectrum  has the maximum height and it is centered around the 
frequency $\o_0\pm\d\o_0$. \\
- Low-frequency noise  strongly affects the line shape of the two 
peaks,  leading to an inhomogeneous broadening.  In comparison, the 
natural laser 
linewidth due to spontaneous emission is negligible.\\
- Although we did not produce a quantitative fit to the data, 
we presented a possible 
explanation of the double-peak structure observed in the 
experiment of Ref. \cite{astafiev}. 
We obtained an estimate of the linewidth due to low frequency noise 
in qualitative agreement with the experiments, and we evaluated the 
locking threshold in the injection locking experiment. 

We acknowledge fruitful discussions with O. Astafiev, A. Fedorov and  
M. Marthaler. The work is part of the EU IST
Project EUROSQIP\\

\appendix
\section{}
\label{appendix}

Here we briefly discuss the validity of the model introduced  in 
Section \ref{model}, when applied to describe the experiment of Astafiev \emph{et 
al.} \cite{astafiev}.  The single-qubit laser realized in  Ref. 
\cite{astafiev}, consists of  a  charge qubit coupled capacitively 
to a single-mode electrical resonator and can be thus described 
in the qubit's eigenbasis by the  Hamiltonian 
\be
   H = \hf \D E \s_z + \hbar \o_0 a^{\dagger} a - \hbar g_0 \lf( \sin\zeta\s_z + \cos\zeta\s_x \rg)  \lf( a + a^{\dagger} \rg). \label{DiagHamil}
\ee 
The angle $\zeta$ and the qubit energy splitting depend on the 
charging and Josephson energies, $\ve_{\rm ch}$ and $E_J$, 
$\tan\zeta = \frac{\ve_{ch}}{E_J}$ and $\D E = \sqrt{\ve_{ch}^2+E_J^2}$. %
In order to identify the one- and two-photon couplings, we  now apply 
a Schrieffer-Wolff transformation $U = e^{iS}$ with $S=i \frac{g_0 
\sin \zeta}{\o_0} \s_z \lf( a - a^{\dagger} \rg)$ and perform a 
perturbation expansion in the parameter $g_0/\o_0$. The transformed Hamiltonian, 
$\tilde H= U^{\dag}HU$, thus becomes
\bea \tilde H & \simeq & \hf \Delta E \sigma_z + \hbar \omega_0 
a^{\dagger} a + \hbar g_1 \s_x \lf( a + a^{\dagger} \rg) \\ 
\nonumber & & + \hbar g_2 i \s_y  \lf( a^2 - \lf( a^{\dagger} \rg)^2 
\rg) . \eea
Here we neglected terms of order $\lf(g_0/\o_0\rg)^3$ and  
introduced the two coupling constants
 $g_1 = - g_0 \cos \zeta$ and $g_2 = -\frac{2 g_0^2}{\o_0} \sin \zeta \cos \zeta $ 
 for one-photon and two-photon transitions, respectively. 
For the parameters used in the experiment the coupling $g_2$ is 
roughly two orders of magnitude smaller than the one photon coupling 
and  below the semiclassical threshold for the two-photon lasing, $ 
g_2^{\rm thr}=\sqrt{\k^2\G_\vf/(\G_1 D_0^2)}$ \cite{wang}. In the 
parameters regime explored in the experiments, the Hamiltonian  used 
in Eq. (\ref{H}) gives thus a good description of the dynamics of 
the system.

\end{document}